\documentclass[12pt]{article}
\usepackage{amsmath}
\usepackage{color}
\usepackage{amssymb}
\usepackage{bm}

\makeatletter\@addtoreset{equation}{section}
\makeatother
\allowdisplaybreaks[1]

\begin{document}
\begin{titlepage}

\begin{flushright}
\phantom{preprint no.}
\end{flushright}
\vspace{0.5cm}
\begin{center}
{\Large \bf
On uniqueness of non-vacuum stationary axisymmetric type D spacetime
\vspace{2mm}
}
\lineskip .75em
\vskip0.5cm
{\large Hiroaki Nakajima${}^{1}$, Ya Guo${}^{2}$ and Wenbin Lin${}^{1,\,3,\,*}$}
\vskip 2.5em
${}^{1}$ {\normalsize\it \normalsize\it School of Mathematics and Physics, 
University of South China, \\Hengyang, 421001, China\\
}
\vskip 1.0em
${}^{2}$ {\normalsize\it Shandong Key Laboratory of Space Environment and Exploration Technology, \\
College of Physics and Electronic information, Dezhou University, \\
Dezhou, 253023, China\\}
\vskip 1.0em
${}^{3}$ {\normalsize\it School of Physical Science and Technology, Southwest Jiaotong University, \\ Chengdu, 610031, China\\}
\vskip 1.0em
${}^{*}$ {\normalsize\it Email: lwb@usc.edu.cn\\}
\vskip 1.0em
\vskip 3.0em
\end{center}
\begin{abstract}
We study a family of the non-vacuum stationary axisymmetric type D metrics, where the two principal null directions are geodesic and shearfree. 
It is demonstrated that the conformal-to-Carter metric gives the most general form of this family, without any assumptions which are previously used by  Ovcharenko and Podolsk\'{y} to arrive at this conclusion. 

\end{abstract}
\end{titlepage}

\section{Introduction}

Study of the Petrov type D spacetime \cite{Petrov} is intriguing, which contains several important black hole spacetime, 
such as the Kerr-Newman black holes and its special cases (Schwarzscild, Reissner-Nordstr\"{o}m, Kerr, etc.). 
The type D metric has the two doubly-degenerated principal null directions (PNDs),  
which are shown to be both geodesic and shearfree in the vacuum. 
Conversely, if the two real null tetrads are both geodesic and shearfree in the type D, then they are the PNDs. 
It is known as the Goldberg-Sachs (GS) theorem \cite{GS}.
Moreover, in the vacuum, the type D metric is completely classified \cite{Kinnersley:1969zza}. 
There is also the extensions of the GS theorem to the case of some non-vacuum backgrounds \cite{KT0,KT,RS,Stephani:2003tm}, 
such as the case of the electromagnetic (EM) field aligned to both of the PNDs in the Einstein-Maxwell system \cite{KT0}.  
In this case, the exact solution is shown to be given by the Pleba\'{n}ski-Demia\'{n}ski (PD) metric \cite{Plebanski:1976gy}, 
which can express all the known solutions in this class by appropriate limiting procedure 
\cite{Griffiths:2005se,Griffiths:2005qp,Podolsky:2022xxd,Astorino:2023uim,Astorino:2024bfl,Ovcharenko:2024yyu,Wu:2024tuh,Podolsky:2026syr}. 
In the general non-vacuum type D spacetime, the generalized GS theorem does not necessarily  hold, and then the condition that the two PNDs 
are both geodesic and shearfree is separately assumed if needed. 

On the other hand, in the case of the non-aligned EM field, the study of the exact solutions has developed rather recently. 
Van den Bergh and Carminati have studied the most general static (non-twisting) type D spacetime \cite{VandenBergh:2020lvf}
under the assumption that the PNDs are geodesic and shearfree. 
Ovcharenko and Podolsk\'{y} (OP) have extended the above by including the twist, 
and have found a stationary axisymmetric type D metric, referred to as the OP metric \cite{Podolsky:2025tle,Ovcharenko:2025cpm}, 
which can be thought as the Kerr black hole in the background of the Bertotti-Robinson magnetic field \cite{GP}. 
They have also obtained an important result \cite{Ovcharenko:2026pow} about the uniqueness, where if a spacetime satisfies the following conditions:  
\begin{enumerate}
\item[i)] It is stationary and axisymmetric with non-null Killing vectors. 
\item[ii)] It is of the Petrov type D. 
\item[iii)] The two PNDs are geodesic and shearfree. 
\item[iv)] The two PNDs are orthogonal to the \textit{polar} direction given in \cite{Ovcharenko:2026pow}. 
\item[v)] A specific one-form is closed. 
\end{enumerate}
Then the metric of the spacetime is claimed to be the (off-shell) conformal-to-Carter one~\cite{Carter:1968ks,Gray:2025lwy}, which is also referred to as the generalized PD metric~\cite{nakajima2025}. 
Note that this result does not rely on the equation of motion, such as the Einstein(-Maxwell) equation and its modified versions.  
Moreover, they have also shown that in the Einstein-Maxwell system, the PD metric for the doubly aligned EM field 
and the OP metric for the non-aligned EM field are the most general forms under the above assumptions. 

One of the extra assumption iv) is introduced in \cite{Ovcharenko:2026pow}, which makes the computation simpler, 
and the other assumption v) is used in \cite{DM,Debever:1984yxe}, where the authors consider the solution in the Einstein-Maxwell system, 
but did not assume the stationarity and the axial symmetry. 
On the other hand, in our previous paper \cite{nakajima2025}, 
we have shown that the assumptions i), ii) and iii) are enough to obtain the conformal-to-Carter metric in a specific frame of the tetrads,
which is equivalent to assume that in a given tetrad basis, two real null tetrads can be regarded as the PNDs.  
In this paper, we will consider the transformation of the frame and will show that the result of our previous paper is available for any frames. 

The remainder of this paper is organized as follows: in section \ref{rot}, we will give the tetrad rotation for finding the PNDs. 
Actually this is rather the well-known procedure.  However, in order to give the explicit relation between the original tetrads and the rotated ones  
(see \eqref{rotation3} below), we will explain it here. 
In section \ref{met}, we will apply the obtained rotation to the tetrads from the general stationary axisymmetric type D metric. 
We then show that the assumption about the frame of the tetrads in our previous paper \cite{nakajima2025} can be removed.  
In section \ref{D}, we will give the summary of our previous paper on the part of obtaining the conformal-to-Carter metric 
for self-containedness.  
Section \ref{summary} is devoted to conclusion and discussion.

\section{Tetrad rotation for type D spacetime}\label{rot}

We consider the null tetrads $l$, $n$, $m$ and $\bar{m}$ in the Newman-Penrose formalism \cite{Newman:1961qr} 
as the one-form basis, such as $l=l_{\mu}dx^{\mu}$, etc, 
which satisfies the orthonormality conditions 
\begin{gather}
l_{\mu}n^{\mu}=-m_{\mu}\bar{m}^{\mu}=1,\quad (\text{other inner products})=0,  
\label{ortho} 
\end{gather}
where we follow the notation of \cite{Newman:1961qr,Pirani}.  
We require that the spacetime should be of the Petrov type D. 
However the real tetrads $l$ and $n$ are not necessarily the doubly degenerated PNDs. 
In order to find the PNDs, we will use the tetrad rotation (local Lorentz transformation) \cite{Janis:1965tx,Chandrasekhar:1985kt}. 
From the type D condition, the following quartic equation 
\begin{gather}
f(z)=\Psi_{4}z^{4}+4\Psi_{3}z^{3}+6\Psi_{2}z^{2}+4\Psi_{1}z+\Psi_{0}=0, 
\label{quartic1}
\end{gather}
has two distinct double roots $z_{+}$ and $z_{-}$, 
where $\Psi_{0}$, $\Psi_{1}$, $\Psi_{2}$, $\Psi_{3}$ and $\Psi_{4}$ are the Weyl scalars. 
Now we first take the tetrad rotation of the second kind with the rotation parameter 
$\mathfrak{b}=z_{-}$ as 
\begin{align}
\hat{l}&=l+\bar{z}_{-}m+z_{-}\bar{m}+z_{-}\bar{z}_{-}n,
\notag\\
\hat{n}&=n, 
\notag\\
\hat{m}&=m+z_{-}n, 
\notag\\
\bar{\hat{m}}&=\bar{m}+\bar{z}_{-}n, 
\label{rotation1}
\end{align}
under which the Weyl scalars are transformed as 
\begin{align}
\hat{\Psi}_{0}&=\Psi_{0}+4z_{-}\Psi_{1}+6z_{-}^{2}\Psi_{2}+4z_{-}^{3}\Psi_{3}+z_{-}^{4}\Psi_{4}=0, 
\notag\\
\hat{\Psi}_{1}&=\Psi_{1}+3z_{-}\Psi_{2}+3z_{-}^{2}\Psi_{3}+z_{-}^{3}\Psi_{4}=0, 
\notag\\
\hat{\Psi}_{2}&=\Psi_{2}+2z_{-}\Psi_{3}+z_{-}^{2}\Psi_{4}, 
\notag\\
\hat{\Psi}_{3}&=\Psi_{3}+z_{-}\Psi_{4}, 
\notag\\
\hat{\Psi}_{4}&=\Psi_{4}.  
\label{weyl1}
\end{align}
Here we have used $\hat{\Psi}_{0}=\hat{\Psi}_{1}=0$ from the assumption that $z_{-}$ is one of the double root of \eqref{quartic1}, 
which implies that $\hat{l}$ is one of the PNDs. The quartic equation \eqref{quartic1} itself is transformed as 
\begin{gather}
f(z)=\hat{\Psi}_{4}\hat{z}^{4}+4\hat{\Psi}_{3}\hat{z}^{3}+6\hat{\Psi}_{2}\hat{z}^{2}=0, 
\label{quartic2}
\end{gather}
where $\hat{z}=z-z_{-}$, and \eqref{quartic2} has two distinct double roots $\hat{z}=0,\,z_{+}-z_{-}$. 

Next we will take the tetrad rotation of the first kind with the parameter $\mathfrak{a}=(\bar{z}_{+}-\bar{z}_{-})^{-1}$ as 
\begin{align}
\check{l}&=\hat{l}
\notag\\
&=l+\bar{z}_{-}m+z_{-}\bar{m}+z_{-}\bar{z}_{-}n,
\notag\\
\check{n}&=\hat{n}+\bar{\mathfrak{a}}\hat{m}+\mathfrak{a}\bar{\hat{m}}+\mathfrak{a}\bar{\mathfrak{a}}\hat{l} 
\notag\\
&=\frac{1}{|z_{+}-z_{-}|^{2}}(l+\bar{z}_{+}m+z_{+}\bar{m}+z_{+}\bar{z}_{+}n),
\notag\\
\check{m}&=\hat{m}+\mathfrak{a}\hat{l} 
\notag\\
&=\frac{1}{\bar{z}_{+}-\bar{z}_{-}}(l+\bar{z}_{+}m+z_{-}\bar{m}+z_{-}\bar{z}_{+}n),
\notag\\
\bar{\check{m}}&=\bar{\hat{m}}+\bar{\mathfrak{a}}\hat{l} 
\notag\\
&=\frac{1}{z_{+}-z_{-}}(l+\bar{z}_{-}m+z_{+}\bar{m}+z_{+}\bar{z}_{-}n). 
\label{rotation2}
\end{align}
The Weyl scalars are transformed as 
\begin{align}
\check{\Psi}_{0}&=\hat{\Psi}_{0}=0, 
\notag\\
\check{\Psi}_{1}&=\hat{\Psi}_{1}+\bar{\mathfrak{a}}\hat{\Psi}_{0}=0, 
\notag\\
\check{\Psi}_{2}&=\hat{\Psi}_{2}+2\bar{\mathfrak{a}}\hat{\Psi}_{1}+\bar{\mathfrak{a}}^{2}\hat{\Psi}_{0}=\hat{\Psi}_{2}, 
\notag\\
\check{\Psi}_{3}&=\hat{\Psi}_{3}+3\bar{\mathfrak{a}}\hat{\Psi}_{2}+3\bar{\mathfrak{a}}^{2}\hat{\Psi}_{1}+\bar{\mathfrak{a}}^{3}\hat{\Psi}_{0}
=\hat{\Psi}_{3}+3\bar{\mathfrak{a}}\hat{\Psi}_{2}=0, 
\notag\\
\check{\Psi}_{4}&=\hat{\Psi}_{4}+4\bar{\mathfrak{a}}\hat{\Psi}_{3}+6\bar{\mathfrak{a}}^{2}\hat{\Psi}_{2}
+4\bar{\mathfrak{a}}^{3}\hat{\Psi}_{1}+\bar{\mathfrak{a}}^{4}\hat{\Psi}_{0}
=\hat{\Psi}_{4}+4\bar{\mathfrak{a}}\hat{\Psi}_{3}+6\bar{\mathfrak{a}}^{2}\hat{\Psi}_{2}=0, 
\label{weyl2}
\end{align}
where we have used that from \eqref{quartic2}, $\bar{\mathfrak{a}}=(z_{+}-z_{-})^{-1}$ is the double root of 
$\hat{\Psi}_{4}+4\bar{\mathfrak{a}}\hat{\Psi}_{3}+6\bar{\mathfrak{a}}^{2}\hat{\Psi}_{2}$=0. 
Now we have $\check{\Psi}_{0}=\check{\Psi}_{1}=\check{\Psi}_{3}=\check{\Psi}_{4}=0$ as \eqref{weyl2}, 
which implies that the real null tetrads $\check{l}$ and $\check{n}$ are both the PNDs. 

For later convenience, we change the normalization of the null tetrads using the diagonal rotation (the third kind) as 
\begin{align}
\tilde{l}&=\frac{1}{|1-z_{+}^{-1}z_{-}|}\check{l}
=\frac{1}{|1-z_{+}^{-1}z_{-}|}(l+\bar{z}_{-}m+z_{-}\bar{m}+z_{-}\bar{z}_{-}n), 
\notag\\
\tilde{n}&=|1-z_{+}^{-1}z_{-}|\check{n}
=\frac{1}{|1-z_{+}^{-1}z_{-}|}(n+\bar{z}_{+}^{-1}\bar{m}+z_{+}^{-1}m+z_{+}^{-1}\bar{z}_{+}^{-1}l),
\notag\\
\tilde{m}&=\frac{1-\bar{z}_{+}^{-1}\bar{z}_{-}}{|1-z_{+}^{-1}z_{-}|}\check{m}
=\frac{1}{|1-z_{+}^{-1}z_{-}|}(m+\bar{z}_{+}^{-1}l+z_{-}n+z_{-}\bar{z}_{+}^{-1}\bar{m}), 
\notag\\
\bar{\tilde{m}}&=\frac{1-z_{+}^{-1}z_{-}}{|1-z_{+}^{-1}z_{-}|}\bar{\check{m}}
=\frac{1}{|1-z_{+}^{-1}z_{-}|}(\bar{m}+\bar{z}_{-}n+z_{+}^{-1}l+z_{+}^{-1}\bar{z}_{-}m). 
\label{rotation3}
\end{align}
Here the real null tetrads $\tilde{l}$ and $\tilde{n}$ are both the PNDs again. For the Weyl tensors, we have 
\begin{gather}
\tilde{\Psi}_{0}=\tilde{\Psi}_{1}=\tilde{\Psi}_{3}=\tilde{\Psi}_{4}=0,\quad \tilde{\Psi}_{2}=\check{\Psi}_{2}. 
\label{weyl0}
\end{gather}
Because we construct the new tetrad basis by the tetrad rotation from the original one, 
it is obvious that the null tetrads $(\tilde{l},\,\tilde{n},\,\tilde{m},\,\bar{\tilde{m}})$ satisfy the orthonormality conditions 
\begin{gather}
\tilde{l}_{\mu}\tilde{n}^{\mu}=-\tilde{m}_{\mu}\bar{\tilde{m}}^{\mu}=1,\quad (\text{other inner products})=0,  
\end{gather}
as the original tetrads satisfy \eqref{ortho}. Moreover the two basis give the same metric as 
\begin{gather}
ds^{2}=2ln-2m\bar{m}=2\tilde{l}\tilde{n}-2\tilde{m}\bar{\tilde{m}}. 
\end{gather}

\section{Stationary axisymmetric type D metric}\label{met}

Now we will consider a stationary axisymmetric metric satisfying the Petrov type D condition with non-null Killing vectors,  
using the Newman-Penrose formalism \cite{Newman:1961qr}. 
We begin with the metric 
\begin{gather}
ds^{2}=Sdt^{2}-2Tdtd\varphi+Ud\varphi^{2}-C dr^{2}-D d\theta^{2},
\label{metric0}
\end{gather}
where $S$, $T$, $U$, $C$ and $D$ are the real functions of $r$ and $\theta$.  
\eqref{metric0} is the most general form of the stationary axisymmetric metric. 
In order to have the Lorentzian signature, 
we have $C,\,D,\,T^{2}-SU>0$ under the $(+---)$ notation. 
We also assume that that $S$ and $U$ are in general nonvanishing, which implies that 
the Killing vectors $\partial/\partial t$ and $\partial/\partial\varphi$ are non-null. 
Instead of \eqref{metric0}, hereafter we will use the following form of the metric as 
\begin{gather}
ds^{2}=A(dt-E_{1} d\varphi)^{2}-B(d\varphi-E_{2} dt)^{2}-C dr^{2}-D d\theta^{2},
\label{metric1}
\end{gather}
and the corresponding null tetrads as
\begin{align}
l&=\sqrt{\frac{A}{2}}(dt-E_{1} d\varphi)-\sqrt{\frac{C}{2}}dr,
\notag\\
n&=\sqrt{\frac{A}{2}}(dt-E_{1} d\varphi)+\sqrt{\frac{C}{2}}dr,
\notag\\
m&=\sqrt{\frac{B}{2}}(d\varphi-E_{2} dt)-i\sqrt{\frac{D}{2}}d\theta,
\notag\\
\bar{m}&=\sqrt{\frac{B}{2}}(d\varphi-E_{2} dt)+i\sqrt{\frac{D}{2}}d\theta, 
\label{tetrads1}
\end{align}
where $A$, $B$, $E_{1}$ and $E_{2}$ are the real functions of $r$ and $\theta$. 
One can find that \eqref{metric1} has one more undetermined function than \eqref{metric0}, 
which means that a set of $(A,\,B,\,E_{1},\,E_{2})$ is not unique for fixed $(S,\,T,\,U)$. Indeed, there is a specific tetrad rotation  
parametrized by a real function $\psi(r,\theta)$, acting on $(A,\,B,\,E_{1},\,E_{2})$ as
\begin{align}
\sqrt{A'}&=\sqrt{A}\cosh\psi-\sqrt{B}E_{2}\sinh\psi, 
\notag\\
\sqrt{A'}E'_{1}&=\sqrt{A}E_{1}\cosh\psi-\sqrt{B}\sinh\psi, 
\notag\\
\sqrt{B'}&=\sqrt{B}\cosh\psi-\sqrt{A}E_{1}\sinh\psi, 
\notag\\
\sqrt{B'}E'_{2}&=\sqrt{B}E_{2}\cosh\psi-\sqrt{A}\sinh\psi.  
\end{align}
This transformation keeps $S$, $T$ and $U$ invariant as 
\begin{align}
S&=A-BE_{2}^{2}=A'-B'(E'_{2})^{2}, 
\notag\\
T&=AE_{1}-BE_{2}=A'E'_{1}-B'E'_{2}, 
\notag\\
U&=AE_{1}^{2}-B=A'(E'_{1})^{2}-B'. 
\end{align}
By including $\psi$, we regard $A$, $B$, $E_{1}$ and $E_{2}$ as the independent degrees of freedom. 
The condition $T^{2}-SU>0$ can be satisfied by requiring
\begin{gather}
AB>0, \quad 1-E_{1}E_{2}\neq 0, 
\end{gather}
and in particular, we take $A$ and $B$ to be positive, which has been used in \eqref{metric1}. 
If $A$ and $B$ are both negative, then by redefinition   
\begin{gather}
\hat{A}=|B|E_{2}^{2}, \quad \hat{B}=|A|E_{1}^{2}, \quad \hat{E}_{1}=E_{2}^{-1}, \quad \hat{E}_{2}=E_{1}^{-1}, 
\end{gather}
the metric is again of the form of \eqref{metric1} with positive $\hat{A}$ and $\hat{B}$.  

The metric \eqref{metric0} and \eqref{metric1} are invariant under the discrete symmetry
\begin{gather}
t \to -t,\quad \varphi \to -\varphi,  
\label{reflection}
\end{gather}
where the tetrads \eqref{tetrads1} transform as 
\begin{gather}
l \to -n, \quad n \to -l, \quad m\to -\bar{m}, \quad \bar{m} \to -m. 
\label{reflection2}
\end{gather}
Then the invariance under \eqref{reflection} gives the special relations among the spin coefficients and the Weyl scalars as \cite{Tanatarov:2012gf}  
\begin{gather}
\kappa=\nu, \quad \sigma=\lambda, \quad \tau=\pi, \quad \rho=\mu, \quad \epsilon=\gamma, \quad \alpha=\beta, 
\label{spin}
\\
\Psi_{0}=\Psi_{4}, \quad \Psi_{1}=\Psi_{3}.   
\label{weyl}
\end{gather}
From \eqref{weyl}, the quartic equation \eqref{quartic1} becomes a reciprocal (palindromic) equation, 
and hence by introducing $w=z+z^{-1}$, \eqref{quartic1} can be decomposed as 
\begin{align}
\Psi_{0}w^{2}+4\Psi_{1}w+6\Psi_{2}-2\Psi_{0}&=0, 
\label{quad1}
\\
z^{2}-wz+1&=0, 
\label{quad2} 
\end{align}
where the type D condition requires that either \eqref{quad1} or \eqref{quad2} should have the double root, but not both. 
The authors of \cite{Tanatarov:2012gf} referred to the case that \eqref{quad1} has the double root as \textit{generic}, 
and the case that \eqref{quad2} has the double root is separately considered. 
For the generic case, we have the vanishing discriminant of \eqref{quad1} as 
\begin{gather}
\Psi_{0}^{2}+2\Psi_{1}^{2}-3\Psi_{0}\Psi_{2}=0, 
\label{case1}
\end{gather}
and $w=-2\Psi_{1}/\Psi_{0}$. Then $z_{\pm}$ are given by 
\begin{gather}
z_{\pm}=-\frac{\Psi_{1}}{\Psi_{0}}\pm\sqrt{\frac{\Psi_{1}^{2}}{\Psi_{0}^{2}}-1}\,, 
\label{root1}
\end{gather}
and in particular, 
\begin{gather}
z_{+}z_{-}=1. 
\label{z}
\end{gather}
On the other hand, for the non-generic case, from the assumption that \eqref{quad2} has the double root, we have $w=\pm 2$, 
which should be the roots of \eqref{quad1}.  
This implies
\begin{gather}
\Psi_{1}=0, \quad \Psi_{0}+3\Psi_{2}=0, 
\end{gather}
and we can take 
\begin{gather}
z_{+}=1,\quad z_{-}=-1. 
\label{root2}
\end{gather}
Because the important examples of the type D spacetime, such as the Kerr-Newman black hole and its special cases, 
are all generic, hereafter we focus on the generic type D spacetime. 

The tetrad basis \eqref{rotation3} with two PNDs are computed from \eqref{tetrads1} as 
\begin{align}
\tilde{l}&=\frac{1}{\sqrt{2}}\left[\sqrt{\tilde{A}}(dt-\tilde{E}_{1}d\varphi)
-\frac{1-z_{-}\bar{z}_{-}}{|1-z_{-}^{2}|}\sqrt{C}dr+\frac{i(z_{-}-\bar{z}_{-})}{|1-z_{-}^{2}|}\sqrt{D}d\theta\right], 
\notag\\[2mm]
\tilde{n}&=\frac{1}{\sqrt{2}}\left[\sqrt{\tilde{A}}(dt-\tilde{E}_{1}d\varphi)
+\frac{1-z_{-}\bar{z}_{-}}{|1-z_{-}^{2}|}\sqrt{C}dr-\frac{i(z_{-}-\bar{z}_{-})}{|1-z_{-}^{2}|}\sqrt{D}d\theta\right],
\notag\\[2mm]
\tilde{m}&=\frac{1}{\sqrt{2}}\left[\sqrt{\tilde{B}}(d\varphi-\tilde{E}_{2}dt)
+\frac{(z_{-}-\bar{z}_{-})}{|1-z_{-}^{2}|}\sqrt{C}dr-i\frac{1-z_{-}\bar{z}_{-}}{|1-z_{-}^{2}|}\sqrt{D}d\theta\right],
\notag\\[2mm]
\bar{\tilde{m}}&=\frac{1}{\sqrt{2}}\left[\sqrt{\tilde{B}}(d\varphi-\tilde{E}_{2}dt)
-\frac{(z_{-}-\bar{z}_{-})}{|1-z_{-}^{2}|}\sqrt{C}dr+i\frac{1-z_{-}\bar{z}_{-}}{|1-z_{-}^{2}|}\sqrt{D}d\theta\right],
\label{tetrads2}
\end{align}
where we have used \eqref{z} and have introduced the functions $\tilde{A}$, $\tilde{B}$, $\tilde{E}_{1}$ and $\tilde{E}_{2}$ as 
\begin{align}
\sqrt{\tilde{A}}&=\frac{1+z_{-}\bar{z}_{-}}{|1-z_{-}^{2}|}\sqrt{A}-\frac{z_{-}+\bar{z}_{-}}{|1-z_{-}^{2}|}\sqrt{B}E_{2}, 
\notag\\
\sqrt{\tilde{A}}\tilde{E}_{1}&=\frac{1+z_{-}\bar{z}_{-}}{|1-z_{-}^{2}|}\sqrt{A}E_{1}-\frac{z_{-}+\bar{z}_{-}}{|1-z_{-}^{2}|}\sqrt{B},
\notag\\
\sqrt{\tilde{B}}&=\frac{1+z_{-}\bar{z}_{-}}{|1-z_{-}^{2}|}\sqrt{B}-\frac{z_{-}+\bar{z}_{-}}{|1-z_{-}^{2}|}\sqrt{A}E_{1}, 
\notag\\
\sqrt{\tilde{B}}\tilde{E}_{2}&=\frac{1+z_{-}\bar{z}_{-}}{|1-z_{-}^{2}|}\sqrt{B}E_{2}-\frac{z_{-}+\bar{z}_{-}}{|1-z_{-}^{2}|}\sqrt{A}. 
\end{align}
We also introduce the new coordinates $\tilde{r}$ and $\tilde{\theta}$, and the functions $\tilde{C}$ and $\tilde{D}$ by the following relation as 
\begin{gather}
\begin{pmatrix} \sqrt{\tilde{C}}d\tilde{r} \\[5mm] \sqrt{\tilde{D}}d\tilde{\theta} \end{pmatrix}=
\begin{pmatrix} 
\displaystyle{\frac{1-z_{-}\bar{z}_{-}}{|1-z_{-}^{2}|}} & \displaystyle{-\frac{i(z_{-}-\bar{z}_{-})}{|1-z_{-}^{2}|}}
\\[5mm]
\displaystyle{\frac{i(z_{-}-\bar{z}_{-})}{|1-z_{-}^{2}|}} & \displaystyle{\frac{1-z_{-}\bar{z}_{-}}{|1-z_{-}^{2}|}}
\end{pmatrix}
\begin{pmatrix} \sqrt{C}dr \\[5mm] \sqrt{D}d\theta \end{pmatrix}. 
\end{gather}
The existence of $\tilde{r}$ and $\tilde{\theta}$ as the functions of $r$ and $\theta$ is guaranteed from the Frobenius theorem 
without any other conditions, because we only have two variables. 
The linear independence between $\tilde{r}$ and $\tilde{\theta}$ can be checked by computing the determinant of the above matrix as 
\begin{gather}
\det
\begin{pmatrix} 
\displaystyle{\frac{1-z_{-}\bar{z}_{-}}{|1-z_{-}^{2}|}} & \displaystyle{-\frac{i(z_{-}-\bar{z}_{-})}{|1-z_{-}^{2}|}}
\\[5mm]
\displaystyle{\frac{i(z_{-}-\bar{z}_{-})}{|1-z_{-}^{2}|}} & \displaystyle{\frac{1-z_{-}\bar{z}_{-}}{|1-z_{-}^{2}|}}
\end{pmatrix}
=1. 
\end{gather}
Using these new coordinates and functions, the new tetrad basis can be expressed as 
\begin{align}
\tilde{l}&=\sqrt{\frac{\tilde{A}}{2}}(dt-\tilde{E}_{1}d\varphi)-\sqrt{\frac{\tilde{C}}{2}}d\tilde{r},  
\notag\\[2mm]
\tilde{n}&=\sqrt{\frac{\tilde{A}}{2}}(dt-\tilde{E}_{1}d\varphi)+\sqrt{\frac{\tilde{C}}{2}}d\tilde{r}, 
\notag\\[2mm]
\tilde{m}&=\sqrt{\frac{\tilde{B}}{2}}(d\varphi-\tilde{E}_{2}dt)-i\sqrt{\frac{\tilde{D}}{2}}d\tilde{\theta}, 
\notag\\[2mm]
\bar{\tilde{m}}&=\sqrt{\frac{\tilde{B}}{2}}(d\varphi-\tilde{E}_{2}dt)+i\sqrt{\frac{\tilde{D}}{2}}d\tilde{\theta}, 
\label{tetrads3}
\end{align}
which has the same form as the original basis \eqref{tetrads1} up to the replacement of the coordinates and the functions as 
\begin{gather}
(r,\,\theta)\to(\tilde{r},\,\tilde{\theta}),\quad 
(A,\,B,\,C,\,D,\,E_{1},\,E_{2})\to(\tilde{A},\,\tilde{B},\,\tilde{C},\,\tilde{D},\,\tilde{E}_{1},\,\tilde{E}_{2}). 
\end{gather}
Moreover in the basis \eqref{tetrads3}, the Weyl scalars satisfy \eqref{weyl0} and the spin coefficients satisfy the similar relation with 
\eqref{spin} as 
\begin{gather}
\tilde{\kappa}=\tilde{\nu}, \quad \tilde{\sigma}=\tilde{\lambda}, \quad \tilde{\tau}=\tilde{\pi}, \quad 
\tilde{\rho}=\tilde{\mu}, \quad \tilde{\epsilon}=\tilde{\gamma}, \quad \tilde{\alpha}=\tilde{\beta}. 
\label{spintilde}
\end{gather}
These properties imply that without loss of generality, we can assume that the real null tetrads $l$ and $n$ are the PNDs, and can  
directly apply the condition that those are geodesic and shearfree. 
This leads the conformal-to-Carter metric \cite{Carter:1968ks,Gray:2025lwy} as we have obtained in \cite{nakajima2025} 
(see also next section).

\section{Conformal-to-Carter metric}\label{D}

As we have seen in the previous section, the real null tetrads $l$ and $n$ in \eqref{tetrads1} can be assumed to be the PNDs  
without loss of generality. We will give the brief summary for obtaining the conformal-to-Carter metric by requiring 
$\kappa=\sigma=\nu=\lambda=0$ and $\Psi_{0}=\Psi_{1}=\Psi_{3}=\Psi_{4}=0$ to \eqref{tetrads1} (or \eqref{tetrads3}) explicitly. For more detail, 
see \cite{nakajima2025}%
\footnote{Here the null tetrads $m$ and $\bar{m}$ are different from those in \cite{nakajima2025} by the constant phase rotation.}. 

First the spin coefficients $\kappa=\nu$ and $\sigma=\lambda$ are computed as 
\begin{align}
\kappa=\nu&=\frac{i}{4\sqrt{2D}}\left[
\frac{\partial_{\theta}C}{C}-\frac{\partial_{\theta}A}{A}+2\frac{E_{2}\partial_{\theta}E_{1}}{1-E_{1}E_{2}}
-2i\sqrt{\frac{AD}{BC}}\frac{\partial_{r}E_{1}}{1-E_{1}E_{2}}
\right],
\label{kappa}
\\
\sigma=\lambda&=\frac{1}{4\sqrt{2C}}\left[
\frac{\partial_{r}D}{D}-\frac{\partial_{r}B}{B}+2\frac{E_{1}\partial_{r}E_{2}}{1-E_{1}E_{2}}
+2i\sqrt{\frac{BC}{AD}}\frac{\partial_{\theta}E_{2}}{1-E_{1}E_{2}}
\right].
\label{sigma}
\end{align}
Then $\kappa=\sigma=\nu=\lambda=0$ implies
\begin{align}
&\frac{\partial_{\theta}C}{C}-\frac{\partial_{\theta}A}{A}+2\frac{E_{2}\partial_{\theta}E_{1}}{1-E_{1}E_{2}}=0,
\label{cond1}
\\
&E_{1}=E_{1}(\theta),
\label{cond2}
\\
&\frac{\partial_{r}D}{D}-\frac{\partial_{r}B}{B}+2\frac{E_{1}\partial_{r}E_{2}}{1-E_{1}E_{2}}=0,
\label{cond3}
\\
&E_{2}=E_{2}(r). 
\label{cond4}
\end{align}
From \eqref{cond1} and \eqref{cond4}, we have
\begin{gather}
\partial_{\theta}\left[\frac{C}{A(1-E_{1}E_{2})^{2}}\right]=0,\quad \text{then} \quad 
\frac{C}{A(1-E_{1}E_{2})^{2}}=F(r).
\label{rel1}
\end{gather}
In a similar way, from \eqref{cond2} and \eqref{cond3}, we have
\begin{gather}
\partial_{r}\left[\frac{D}{B(1-E_{1}E_{2})^{2}}\right]=0,\quad \text{then} \quad 
\frac{D}{B(1-E_{1}E_{2})^{2}}=G(\theta).
\label{rel2}
\end{gather}
We will use \eqref{rel1} and \eqref{rel2} to eliminate $C$ and $D$, respectively. 

One can find that $\Psi_{0}=\Psi_{4}=0$ is already satisfied, 
then $\Psi_{1}=\Psi_{3}$ is computed as
\begin{gather}
\Psi_{1}=\Psi_{3}=\frac{-1}{16\sqrt{AB}(1-E_{1}E_{2})^{3}}(I-iJ),
\label{psi1}
\end{gather}
where $I$ and $J$ are given by
\begin{align}
I&=
\frac{2}{F}\partial_{r}\left(\frac{B}{A}E'_{2}(r)\right)+\frac{2}{G}\partial_{\theta}\left(\frac{A}{B}E'_{1}(\theta)\right)
\notag\\
&\qquad {}+(1-E_{1}E_{2})^{2}\frac{B}{A}E'_{2}(r)\partial_{r}\left[\frac{1}{F(1-E_{1}E_{2})^{2}}\right]
\notag\\
&\qquad {}+(1-E_{1}E_{2})^{2}\frac{A}{B}E'_{1}(\theta)\partial_{\theta}\left[\frac{1}{G(1-E_{1}E_{2})^{2}}\right],
\label{I}
\\
J&=\frac{2(1-E_{1}E_{2})}{\sqrt{FG}}\left[
\frac{\partial_{r}\partial_{\theta}B}{B}-\frac{\partial_{r}B\partial_{\theta}B}{B^{2}}
-\frac{\partial_{r}\partial_{\theta}A}{A}+\frac{\partial_{r}A\partial_{\theta}A}{A^{2}}
\right]. 
\label{J}
\end{align}
Here the prime denotes the ordinary derivative of the functions with respect to their arguments.
In order to have $\Psi_{1}=\Psi_{3}=0$, we have to require $I=J=0$. The condition $J=0$ can be integrated as
\begin{gather}
\frac{B(r,\theta)}{A(r,\theta)}=\frac{M(\theta)}{H(r)}. 
\label{rel3}
\end{gather}
By defining
\begin{gather}
P(r)=\frac{1}{F(r)H^{2}(r)},\quad Q(\theta)=\frac{1}{G(\theta)M^{2}(\theta)}, 
\end{gather}
the condition $I=0$ can be rewritten as 
\begin{align}
&\bigl[1-E_{1}(\theta)E_{2}(r)\bigr]\bigl[P'(r)E'_{2}(r)+2P(r)E''_{2}(r)+Q'(\theta)E'_{1}(\theta)+2Q(\theta)E''_{1}(\theta)\bigr]
\notag\\
&{}+2E_{1}(\theta)P(r)\bigl[E'_{2}(r)\bigr]^{2}+2E_{2}(r)Q(\theta)\bigl[E'_{1}(\theta)\bigr]^{2}
\notag\\
&=0. 
\label{funceq}
\end{align}
We further rewrite the above by introducing 
\begin{gather}
\mathcal{P}=P(r)\bigl[E'_{2}(r)\bigr]^{2}, \quad \mathcal{Q}=Q(\theta)\bigl[E'_{1}(\theta)\bigr]^{2}.  
\end{gather}
and by regarding $\mathcal{P}$ and $\mathcal{Q}$ as the functions of $E_{2}$ and $E_{1}$, respectively. 
Then \eqref{funceq} can be expressed as 
\begin{gather}
(1-E_{1}E_{2})\bigl(\mathcal{P}'(E_{2})+\mathcal{Q}'(E_{1})\bigr)+2E_{1}\mathcal{P}(E_{2})+2E_{2}\mathcal{Q}(E_{1})=0. 
\label{funceq2}
\end{gather}
By taking the derivative with respect to $E_{1}$ and $E_{2}$, we have 
\begin{gather}
\bigl(E_{2}\mathcal{P}''(E_{2})-\mathcal{P}'(E_{2})\bigr)+\bigl(E_{1}\mathcal{Q}''(E_{1})-\mathcal{Q}'(E_{1})\bigr)=0, 
\end{gather}
which can now be separated as 
\begin{gather}
E_{2}\mathcal{P}''(E_{2})-\mathcal{P}'(E_{2})=2\mathcal{C}, \quad E_{1}\mathcal{Q}''(E_{1})-\mathcal{Q}'(E_{1})=-2\mathcal{C}. 
\label{diffeq1}
\end{gather}
Here $\mathcal{C}$ is constant. From \eqref{diffeq1}, one can find that $\mathcal{P}$ and $\mathcal{Q}$ are at most quadratic polynomials 
of $E_{2}$ and $E_{1}$, respectively. \eqref{funceq2} can thus be solved as  
\begin{gather}
\mathcal{P}(E_{2})=\mathcal{B}-2\mathcal{C}E_{2}-\mathcal{A}E_{2}^{2}, \quad 
\mathcal{Q}(E_{1})=\mathcal{A}+2\mathcal{C}E_{1}-\mathcal{B}E_{1}^{2}, 
\label{PQ}
\end{gather} 
and hence 
\begin{align}
P(r)&=\frac{\mathcal{B}-2\mathcal{C}E_{2}(r)-\mathcal{A}E_{2}^{2}(r)}{\bigl(E'_{2}(r)\bigr)^{2}},  
\label{rel4}
\\
Q(\theta)&=\frac{\mathcal{A}+2\mathcal{C}E_{1}(\theta)-\mathcal{B}E_{1}^{2}(\theta)}{\bigl(E'_{1}(\theta)\bigr)^{2}}, 
\label{rel5} 
\end{align}
where $\mathcal{A}$ and $\mathcal{B}$ are constant.
By substituting \eqref{rel4} and \eqref{rel5},  the metric is obtained as 
\begin{align}
ds^{2}&=A_{1}(r,\theta)\Biggl[\frac{H(r)}{1-E_{1}(\theta)E_{2}(r)}(dt-E_{1}(\theta) d\varphi)^{2}
-\frac{M(\theta)}{1-E_{1}(\theta)E_{2}(r)}(d\varphi-E_{2}(r) dt)^{2}
\notag\\
&\qquad\qquad\qquad
{}-\frac{1-E_{1}(\theta)E_{2}(r)}{H(r)}\frac{dE_{2}(r)^{2}}{\mathcal{B}-2\mathcal{C}E_{2}(r)-\mathcal{A}E_{2}^{2}(r)}
\notag\\
&\qquad\qquad\qquad
-\frac{1-E_{1}(\theta)E_{2}(r)}{M(\theta)}
\frac{dE_{1}(\theta)^{2}}{\mathcal{A}+2\mathcal{C}E_{1}(\theta)-\mathcal{B}E_{1}^{2}(\theta)}
\Biggr], 
\label{metric2}
\end{align}
where 
\begin{gather}
dE_{1}(\theta)=E'_{1}(\theta)d\theta, \quad dE_{2}(r)=E'_{2}(r)dr. 
\label{e12}
\\
A_{1}(r,\theta)=\frac{1-E_{1}(\theta)E_{2}(r)}{H(r)}A(r,\theta). 
\end{gather}
We note that the appearance of \eqref{e12} implies that the choices of $E_{1}(\theta)$ and $E_{2}(r)$ are the gauge degrees of freedom 
associated with the coordinate transformation of $\theta$ and $r$, respectively. 

Instead of $E_{1}$ and $E_{2}$ themselves, we introduce new functions $R$ and $X$ defined as 
\begin{gather}
E_{1}(\theta)=\left(\frac{a}{b-X^{2}(\theta)}+c\right)^{-1}, 
\quad 
E_{2}(r)=\frac{a}{R^{2}(r)+b}+c, 
\label{defRX}
\end{gather}
and will use $R$ and $X$ as the new coordinates. 
Here the parameters $a$, $b$ and $c$ are given by 
\begin{gather}
\mathcal{A}=4b, \quad \mathcal{B}=-4ac-4bc^{2}, \quad \mathcal{C}=-2a-4bc.  
\label{abc}
\end{gather}
In order to rewrite the above metric into a simpler form, we define the new coordinates $\tilde{t}$ and $\tilde{\varphi}$ as,  
\begin{gather}
\tilde{t}=\left(1+\frac{bc}{a}\right)t-\frac{b}{a}\varphi, \quad \tilde{\varphi}=\frac{1}{a}(\varphi -ct). 
\end{gather}
We also define the new conformal factor $\Omega(r,\theta)$ and the new structure functions $L(r)$ and $K(\theta)$ as
\begin{gather}
\Omega^{-2}(r,\theta)=\frac{a}{(a+bc-cX^{2}(\theta))(R^{2}(r)+b)}A_{1}(r,\theta), 
\notag\\
L(r)=(R^{2}(r)+b)^{2}H(r), \quad K(\theta)=(a+bc-cX^{2}(\theta))^{2}M(\theta). 
\end{gather}
In terms of those quantities, the metric \eqref{metric2} becomes 
\begin{align}
ds^{2}&=\Omega^{-2}(R,X)\Biggl[\frac{L(R)}{R^{2}+X^{2}}(d\tilde{t}+X^{2}d\tilde{\varphi})^{2}-\frac{R^{2}+X^{2}}{L(R)}dR^{2}
\notag\\
&\qquad\qquad\qquad
-\frac{R^{2}+X^{2}}{K(X)}dX^{2}-\frac{K(X)}{R^{2}+X^{2}}(R^{2}d\tilde{\varphi}-d\tilde{t})^{2}
\Biggr], 
\label{metric4}
\end{align}
which is nothing but the conformal-to-Carter metric \cite{Carter:1968ks,Gray:2025lwy}. 
Here as the arguments of the conformal factor and the structure functions, 
we have used $R$ and $X$ instead of $r$ and $\theta$, respectively.

\section{Conclusion and discussion}\label{summary}

In this paper, we have studied the stationary axisymmetric type D metric, where the two PNDs are both geodesic and shearfree. 
We find that the conformal-to-Carter metric is the most general form of this family, 
which corresponds to the extension of the result of our previous paper \cite{nakajima2025} by considering the tetrad rotation, 
and generalization of the result by Ovcharenko and Podolsk\'{y} \cite{Ovcharenko:2026pow}.   
The advantage is that our analysis can be applied to any (generic) case. Using the result in \cite{Ovcharenko:2026pow}, 
one can conclude in more general sense that the most general metric satisfying the above conditions in the Einstein-Maxwell system is 
the PD one \cite{Plebanski:1976gy} with the doubly aligned electromagnetic field, 
and the OP one \cite{Podolsky:2025tle,Ovcharenko:2025cpm} with the non-aligned electromagnetic field. 

Now we can find the reason why the two assumptions iv) and v) in introduction can be removed. 
First is the choice of the appropriate tetrad basis \eqref{rotation3} and \eqref{tetrads3}, in particular $\tilde{m}$ and $\bar{\tilde{m}}$, 
and then we can remove the assumption of orthogonality (assumption iv)). 
Second is that we have used the property that we have the functions dependent only on the two variables $r$ and $\theta$ 
from stationarity and the axial symmetry. 
Because of this, we can apply the Frobenius theorem without any other conditions, 
which is important to remove the assumption of the existence of a specific closed one-form (assumption v)). 

It is rather surprising that the conformal-to-Carter metric itself can be obtained just from the geometrical information, 
and without using the equation of motion such as the Einstein(-Maxwell) equation and its modified version.  
It would be interesting to apply this metric as an ansatz to the case of the different matter, 
the modified gravitational theories, etc. For example, the metric would also be useful to study the black-hole-like objects 
\cite{Vigeland:2011ji, Johannsen:2013szh, Carson:2020dez, Yagi:2023eap}, 
such as the boson star and the gravastar (see e.g.\,\cite{Cardoso:2019rvt} for a review). 
The deviation of such objects from the black hole is parametrically introduced in the metric 
as the different form of the structure functions $\Omega(R,X)$, $L(R)$ and $K(X)$ in \eqref{metric4}. 
The extension to the case without restricting that the PNDs are not necessarily geodesic and shearfree \cite{PH,GP0}, 
the case of the null Killing vectors \cite{DM,Debever:1984yxe}, 
and the case where even the stationarity and the axial symmetry are not assumed  \cite{DM,Debever:1984yxe} are intriguing. 

It would also be interesting to study the gravitational-wave (GW) equation on the background of this spacetime. 
Because the metric is of the type D and the PNDs are geodesic and shearfree, the method of the Teukolsky equation \cite{Teukolsky:1973ha} 
can be directly applied as in \cite{Jing:2021ahx,Jing:2022vks,Guo:2023niy,Jing:2023vzq,Guo:2023hdn,Guo:2023wtx,Nakajima:2024qrq,Jing:2025utt}. 
However for non-vacuum case, there is a problem of gauge dependence, where the form of the (Teukolsky-like) GW equation may depend on the 
choice of the gauge \cite{Guo:2023wtx,Nakajima:2024qrq}, although the unknown variables (the fluctuation of $\Psi_{0}$ and $\Psi_{4}$) 
are gauge-invariant. 
Moreover, for the modified theories of gravity, there would be the mixing between the spin-2 modes and the other (spin-1 and spin-0) modes, 
which has to be separated \cite{Blazquez-Salcedo:2016enn,Wagle:2023fwl,Guo:2024bqe}.


\section*{Acknowledgements}
This work was supported in part by the National Natural Science Foundation of China (Grant No. 12475057) 
and Shandong Provincial Natural Science Foundation (Grant No. ZR2026QC0026)

\begin{appendix}

\end{appendix}

\end{document}